\documentclass[twocolumn,showpacs,preprintnumbers,amsmath,amssymb,prb,aps]{revtex4-1}
\usepackage{epsfig}
\usepackage{graphicx}
\usepackage{dcolumn}
\usepackage{bm}
\usepackage{textcomp}

\begin{document}

\title{The Importance of Temperature Dependent Energy Gap in the Understanding of High Temperature Thermoelectric Properties}
\author{Saurabh Singh} 
\altaffiliation{Electronic mail:saurabhsingh950@gmail.com}
\author{Sudhir K. Pandey}
\altaffiliation{Electronic mail:sudhir@iitmandi.ac.in}
\affiliation{School of Engineering, Indian Institute of Technology Mandi, Kamand 175005, Himachal Pradesh, India}

\date{\today}

\begin{abstract}

    In the present work, we show the importance of temperature dependent energy band gap, E$_{g}$(T), in understanding the high temperature thermoelectric (TE) properties of material by considering LaCoO$_{3}$ (LCO) and ZnV$_{2}$O$_{4}$ (ZVO) compounds as a case study. For the fix value of band gap, E$_{g}$, deviation in the values of $\alpha$ has been observed above 360 K and 400 K for LCO and ZVO compounds, respectively. These deviation can be overcomed by consideration of temperature dependent band gap. The change in used value of E$_{g}$ with respect to temperature is $\sim$4 times larger than that of InAs. This large temperature dependence variation in E$_{g}$ can be attributed to decrement in the effective on-site Coulomb interaction due to lattice expansion. At 600 K, the value of \textit{ZT} for \textit{n} and \textit{p-doped}, LCO is $\sim$0.35 which suggest that it can be used as a potential material for TE device. This work clearly suggest that one should consider the temperature dependent band gap in predicting the high temperature TE properties of insulating materials. 
    
\end{abstract}


\maketitle

\section{Introduction} 
  
       In the past few decades, the TE materials are being more explored as an alternate source of energy, since they are suitable to convert the waste heat into useful electricity.\cite{DiSalvo, Bell} The ability of TE materials to efficiently produce the electricity is decided by the \textit{ZT} value, known as \textit{figure-of-merit}.\cite{Snyder} Therefore, in high temperature range applications there are always need of new materials with high value of \textit{ZT}. In the search of new TE materials, experimental and theoretical tools have been extensively used.\cite{Trit, Madsen} The properties obtained by experimental tools are more genuine and utilization of these properties in realistic applications are generally preferred.\cite{Rowe} However, the theoretical tools are advantageous over experimental tools as they are less time consuming and cost effective. In the theoretical tools, density functional theory is very useful in discovery of new materials and prediction of various properties such as electrical conductivity, thermal conductivity, TE power and etc.\cite{Sholl, singhDJ, sonu} In spite of many advantages, DFT tools have certain constraints such as it can only perform the ground state (T= 0 K) calculations of electronic properties. Similarly, in the BoltzTraP code calculations, the temperature dependent properties are done under the constant relaxation time approximations.\cite{Madsen} Therefore, the prediction of the temperature dependent properties of new materials in high temperature region are not reliable under these limitations. Also, at high temperature the various scattering mechanism affect the transport behavior of the material.\cite{Salameh} So, to make a more realistic approach towards this, one should include all the high temperature scattering parameters. The inclusion of these parameters requires consideration of total relaxation time, $\tau $$_{total}$, for which one needs to know about various physical parameters of the material.\cite{Salameh}
    In order to check the reliability of the DFT code for proper understanding of the transport properties of materials, the temperature dependent TE properties of LCO and ZVO compounds has been studied earlier.\cite{Singh, SSingh} The consideration of constant relaxation time ($\tau$) and effective masses of the charge carriers gives a good matching between experimental and theoretically calculated values of $\alpha$ only in the temperature range 300-360 K and 300-400 K for the LCO and ZVO compound, respectively. But, in the high temperature region, i.e. above 400 K, the deviation in the calculated values of $\alpha$ from the experimental values were observed. The maximum deviation in the values of $\alpha$ were $\sim$74 $\mu$V/K at $\sim$600 K for ZVO and $\sim$120$\mu$V/K at $\sim$460K for LCO. This deviation in the $\alpha$ was attributed to the high temperature scattering effect. On the basis of available understanding and proposed model, the large and unphysical values for scattering parameters are required to minimize the deviations of $\alpha$. Apart from these scattering factors, another important parameter of semiconductor is energy gap. In general, the energy gap (E$_{g}$) of the semiconductors are found to be temperature dependent due to change in the inter-atomic spacing.\cite{Ashcroft, Varshni} The values of carriers concentrations (\textit{n}) are decided by E$_{g}$ of the materials. Due to \textit{n} dependency, TE properties of the material can also be understood by consideration of temperature dependent E$_{g}$. This aspects has not been explored much to the best of our knowledge. This gives us a motivation to study the high temperature TE properties by using temperature dependent E$_{g}$.
    In the present work, we have shown the importance of temperature dependent E$_{g}$ in understanding of high temperature TE properties. The suitable values of E$_{g}$ has been adopted for the estimation of TE parameters. The values of $\alpha$ have been estimated in the temperature range 300-600 K, for LCO and ZVO compounds. The calculated values of $\alpha$ matches very well with the experimental values in the entire temperature range. We have also estimated the power factor (PF) and \textit{ZT} values at different temperatures. It is observed that, the lower amount of doping (\textit{p and n-type}) is required for tuning TE properties in these compounds. The obtained values of $\alpha$$^{2}$$\sigma$/$\tau$(power factor with respect to scattering time) and \textit{ZT} gives a more realistic information in the selection of new  TE materials, which can be used in making the high temperature TE devices. This also give a direction to predict new TE materials for high temperature applications, if the temperature dependent behavior of E$_{g}$ is known.     
\section{Experimental and Computational details }   
    The Polycrystalline LaCoO$_{3}$ sample was prepared by using the single step solution combustion method, whereas ZnV$_{2}$O$_{4}$ sample was prepared by using the conventional solid state ceramic route. The step by step synthesis procedure for both the compounds were provided in our earlier work.\cite{Singh, SSingh} The thermopower measurement on pellets having the thickness of 0.5 mm, diameter of $\sim$5 mm for LCO compound; and thickness of $\sim$1 mm, diameter of $\sim$10 mm for ZnVO compound were carried out. The thermopower data for both the compounds were obtained in 300-600 K temperature range by using the home made setup.\cite{Saurabh}\\
    The TE properties of both the compounds were studied by combining the ab-initio calculations implemented in WIEN2k software and transport properties using BoltzTrap code.\cite{Blaha, Madsen} For both the compounds, we have used experimental lattice parameters of LCO and ZVO  reported by Radaelli \textit{et al.}\cite{Radaelli} and Reehuis \textit{et al.},\cite{Reehuis} respectively; and these lattice parameters were widely accepted by the research community. The computational details were provided in our earlier work.\cite{Combineref} Normally it has been seen that LDA calculations overestimate the bond strengths and give underestimated values for lattice parameter (within 2-3\%). The change in lattice parameters by 2-3\% does not give much change in the electronic structure properties. Hence, interpretations of transport properties do not affected much when the experimental lattice parameters are used for \textit{ab-initio} calculations. Although, full structural optimizations are desirable, where one has to  optimize both the lattice parameters as well as atomic positions. The full structural optimizations are normally time consuming process. Therefore, in the present study we directly adopted the experimental lattice parameters for doing the calculations which are capable in explaining the transport properties very well.   
It has been seen that, LSDA calculations underestimate the energy band gap.\cite{Kohn} In case of strongly correlated electron systems, LSDA calculations normally fail to generate the insulating ground state. To overcome this problem, many approximations such as  LSDA+U, GW and MBJ are used.\cite{Perdew,Ansimov,Hedin,Tran} Among these approximations, the LSDA+U scheme is computationally cheaper and has been extensively used to study the strongly correlated electron systems. Therefore, we have carried out LSDA+U calculations for both the compounds and it give the energy band gap closer to the experimentally reported values.\cite{Rogers, Arima} To get the more accurate values of energy band gap one can do GW calculations on these compounds, which is again a time consuming process and also very expensive calculations, particularly if done self-consistently. Due to limitations of computational facilities, the use of GW scheme for estimation of band gap is beyond the scope of present work. It will be interesting to see in the future, whether the temperature dependent values of energy band gap obtained by \textit{ab-initio} calculations are same or not to the values we have used in the understanding of TE properties of both the compounds.
      
\section{Results and Discussion}

    The $\alpha$ versus T plots for LCO and ZVO are shown in Fig. 1a and 1b, respectively. In our earlier work, we have considered the constant energy gap and effective mass of charge carriers in the estimation of $\alpha$ and found the matching between experimental and calculated values only in the 300-360 K and 300-400 K range for LCO and ZVO, respectively.\cite{Combineref} The large deviations in the values of $\alpha$ were observed in the high temperature region. In order to overcome these deviations, temperature dependent values of E$_{g}$ has been considered for both compounds. The temperature dependent values of E$_{g}$ are used in the BoltzTraP code for estimating the TE parameters. Here, in the estimation of $\alpha$ for a given temperature, we have chosen values of E$_{g}$ such that a good match between experimental and calculated values of $\alpha$ are found. The temperature dependent values of E$_{g}$ thus obtained are shown in the inset of Fig 1a and 1b. At 300 K, the used value of E$_{g}$ for LCO is 0.5 eV and a nonlinear decrements of 68\% is observed in 300-500 K range and above this no significant change in Eg is noticed. The variation in E$_{g}$ with T shows resemblance with the $\alpha$ versus T plot of LCO. For ZVO, $\sim$10\% and $\sim$60\% linear decrements are noticed in the E$_{g}$ and corresponding values of $\mid$dEg/dT$\mid$ are $\sim$2$\times$10$^{-4}$ and $\sim$1.3$\times$10$^{-3}$ eV/K for 300-450 K and 450-600 K, respectively. The change in the $dEg/dT$ at $\sim$450 K appears to give an interesting aspect of electronic transition, however further detailed study in this direction is required to establish any such electronic transition. At this point, it is interesting to compare the rate of dE$_{g}$/dT of LCO and ZVO with normal semiconductor which falls into the category of band insulator. In the case of band insulator, the consideration of U$_{eff}$ (on-site coulomb interaction) is not required to generate the energy band gap. Diamond, silicon, germanium, 6H SiC, GaAs, InP and InAs are the band insulating materials and they show the temperature dependent energy band gap.\cite{Varshni} From the reported data by Varshni et al., we have extracted the values of temperature dependent energy band gap of these substances in the 300-600 K temperature range, and found that among all these substances InAs has maximum change in the value of E$_{g}$ with respect to temperature. In 300-600 K, the decrements in E$_{g}$ with respect to temperature is found to be $\sim$26 \% for InAs. In comparison to InAs, the change in E$_{g}$ with respect to temperature is $\sim$2.6 times larger for LCO in 300-600 K, whereas for ZVO it is nearly equal in 300-450 K and $\sim$4 times larger in 450-600 K temperature range. The LCO and ZVO compounds belong to the group of strongly correlated electron system, and the inclusion of on-site coulomb interaction is necessary to generate the gap in these compounds. The energy band gap close to experimentally observed value is found by taking the U$_{eff}$ equal to 2.75 eV and 3.7 eV in \textit{3d} electrons of Co and V atoms for LCO and ZVO, respectively. The U$_{eff}$ value depends on the neighbouring environment of V and Co atoms such as bond lengths and bond angles. The above calculated band gap is corresponding to the ground state, however the increase in temperature may decrease the value of U$_{eff}$ due to lattice expansion resulting in reduction of band gap. Thus, at high temperature the lower U$_{eff}$ in addition to weak electron-ion interaction may give rise to a large decrement in E$_{g}$. However, to confirm this conjecture a detailed study in this direction is required. Thus, at high temperature the lower U$_{eff}$ in addition to weak electron-ion interaction will give rise to a large decrement in E$_{g}$ as U$_{eff}$ is expected to decrease because of lattice expansion. Therefore, in comparison to band insulators, the change in E$_{g}$ with respect to temperature is expected to be more for both the compounds. The good match between experimental and calculated values of $\alpha$ suggests that consideration of temperature dependent gap appears to be more effective in the prediction of realistic values of $\alpha$ in high temperature range.
    In order to predict the TE properties in high temperature region, the various TE parameters at different temperature have been estimated. To obtain the the power factor with respect to scattering time ($\alpha$$^{2}$$\sigma$/$\tau$), we have taken into account the temperature dependent energy gap behavior. The plot of $\alpha$$^{2}$$\sigma$/$\tau$ is more informative as it gives a more practical value of power factor ($\alpha$$^{2}$$\sigma$) if the experimental value of scattering time, $\tau$ is known. At different temperature, the plots of $\alpha$$^{2}$$\sigma$/$\tau$ versus $\mu$ for LCO and ZVO, are shown in Fig 2a and 2b, respectively. The value of $\mu$ = 0 stands for middle of the energy gap, whereas positive and negative values of $\mu$ represents the electron and hole-type doping, respectively. Two peaks in $\alpha$$^{2}$$\sigma$/$\tau$ plot are observed for each temperature, one for positive and another for negative $\mu$. For both compounds, the peak values of $\alpha$$^{2}$$\sigma$/$\tau$ corresponding to each temperature are larger in \textit{p}-doped than that of \textit{n}-doped compound. As the temperature increases from 300 K to 600 K, the peak values of $\alpha$$^{2}$$\sigma$/$\tau$ gets shift towards $\mu$ = 0 eV for both compounds. At 300 K, for \textit{p}-doped LCO the peak value of $\alpha$$^{2}$$\sigma$/$\tau$ at 300 K is found at $\sim$ -467 meV and with increase in temperature it shifts to $\mu$ equal to $\sim$ -273 meV for 600 K, whereas for \textit{n}-doped the peak value of $\alpha$$^{2}$$\sigma$/$\tau$ corresponding to 300 K is obtained at $\sim$295 meV and it shifted to the $\sim$ 144 meV for 600 K.  Similarly, the shift in the $\mu$ values towards middle of the gap are observed for \textit{n} and \textit{p}-doped ZVO. For both \textit{p} and \textit{n}-doped LCO and ZVO, the variations in $\mu$ (corresponding to the peak values of $\alpha$$^{2}$$\sigma$/$\tau$) and $\alpha$$^{2}$$\sigma$/$\tau$ with respect to temperature are shown in Fig. 3a and 3b, respectively.  The plots shown by symbols with solid and dotted lines are corresponding to fix E$_{g}$ and temperature dependent E$_{g}$, respectively. For fix value of E$_{g}$, the values of $\mu$ corresponding to the peak value of $\alpha$$^{2}$$\sigma$/$\tau$ changes almost linearly for \textit{n} and \textit{p-type} doped LCO and ZVO. However, the nonlinear shifts in chemical potential ( corresponding to peak value of $\alpha$$^{2}$$\sigma$/$\tau$) towards $\mu$ = 0 eV is found in doped compounds. The shift in $\mu$ towards lower magnitude gives a important signature of lower amount of doping. The small amount of doping is more preferable. This information will give a guideline to the experimental research community for tuning the TE properties of parent compounds with suitable doping. As evident from the Fig. 3b, for \textit{n} and \textit{p}-doped LCO and ZVO the calculated values of $\alpha$$^{2}$$\sigma$/$\tau$ corresponding to a given temperature do not show any significant difference in the temperature range 300-500 K for both fixed E$_{g}$ and temperature dependent E$_{g}$ . However, a slight difference in the $\alpha$$^{2}$$\sigma$/$\tau$ are observed in the 500-600 K. For LCO, the values of $\alpha$$^{2}$$\sigma$/$\tau$ for \textit{p}-doped compound are larger than that of \textit{n}-doped compound by an amount of $\sim$4 and $\sim$2($\times$10$^{14}$ $\mu$W cm$^{-1}$K$^{-2}$S$^{-1}$) at 300 and 600 K, respectively. However, for \textit{p}-doped ZVO its value is $\sim$6 and $\sim$4($\times$10$^{14}$ $\mu$W cm$^{-1}$K$^{-2}$S$^{-1}$) larger than that of \textit{n}-doped at 300 K and 600 K, respectively.
    The maximum thermoelectric efficiency of a TE material is determined by the \textit{ZT} of that material. Therefore, to know the efficient TE ability of doped compound the values of \textit{ZT} at various different temperature has also been estimated in 300-600 K. ZT have been calculated by using the peak value of $\alpha$$^{2}$$\sigma$/$\tau$ (shown in Fig. 3b) and the $\kappa$ values reported in literature.\cite{Ishitsuka, Pillai} In the present case, we have used $\tau$ equal to 10$^{-14}$ s. For \textit{p} and \textit{n-type} doped compound, the variations in \textit{ZT} with respect to temperature is shown in Fig. 3c. For ZVO (\textit{n} and \textit{p-type} doping), we have observed almost linear temperature dependent behavior of \textit{ZT} in the 300-600 K range, whereas for both type doping in LCO compound the value of \textit{ZT} increases almost linearly in the temperature range 300-450 and with further increase in temperature it becomes almost constant up to 600 K. In case of ZVO, the value of \textit{ZT} for \textit{p-type} doping is larger than that of \textit{n-type} doping, which suggests that \textit{hole-type} doping is more appropriate for enhancing the TE properties of this compound in the 300-600 K. The value of \textit{ZT} for \textit{n-type} doped ZVO is $\sim$0.02 at 300 K and it reaches to $\sim$0.09 at 600 K. In case of LCO, the values of \textit{ZT} are $\sim$0.2 and $\sim$0.25 at 300 K for \textit{n} and \textit{p}-doped compound, respectively. In 500-600 K range, the ZT value of \textit{p}-doped LCO is almost constant with magnitude of $\sim$0.36, which is slightly greater in comparison to \textit{n}-doped LCO with magnitude of $\sim$0.34. The TE devices (TE cooler and TE generator) are made by TE module which consist the two dissimilar TE materials (\textit{n-type} and \textit{p-type} semiconductor).  Almost equal value of \textit{ZT} ($\sim$0.35) obtained in \textit{p} and \textit{n}-doped LCO suggests that this compound can be more suitable for making the TE device with appropriate amount of doping.

\section{Conclusions} 
    In conclusion, we have studied the role of temperature dependent energy band gap in understanding of high temperature TE properties of LaCoO$_{3}$ and ZnV$_{2}$O$_{4}$ compounds. The adoption of suitable values of Eg(T) give a good match between experimental and calculated values of $\alpha$ for both the compounds in the 300-600 K temperature range. The value of chemical potential corresponding to peak value of $\alpha$$^{2}$$\sigma$/$\tau$ gets shifted towards middle of the band gap as temperature increases from 300 K to 600 K. This observation implies that lower amount of doping needed for tuning the TE properties of these compounds. The estimated value of \textit{ZT} for \textit{n} and \textit{p}-doped LaCoO$_{3}$ is found to be $\sim$0.35 in 500-600 K range, and it suggest that this material can be used as a TE material. Most of the TE materials are semiconductors, and their TE properties are highly influenced by the energy band gap. Thus, our work clearly emphasize that for proper understanding of high temperature TE properties, the consideration of the temperature dependent gap is necessary. Also, this aspect can play a significant role in the prediction of  high temperature TE properties of insulating materials.

\section{Figure Captions:}

FIG. 1. (Color online) Seebeck coefficient versus temperature plots. Shown are (a) LaCoO$_{3}$ and (b) ZnV$_{2}$O$_{4}$.

FIG. 2. (Color online) Variation of Power factor with respect to scattering time ($\alpha$$^{2}$$\sigma$/$\tau$) with chemical potential($\mu$). Shown are (a) LaCoO$_{3}$ and (b) ZnV$_{2}$O$_{4}$.

FIG. 3. (Color online) Variation of (a) chemical potential with temperature, (b) Peak value of $\alpha$$^{2}$$\sigma$/$\tau$ with temperature, (c) \textit{figure-of-merit, ZT} with temperature; for \textit{n} and \textit{p}-type doped LaCoO$_{3}$ and ZnV$_{2}$O$_{4}$ compound.

\begin{figure}[htbp]
\caption{}
\includegraphics[scale=0.5]{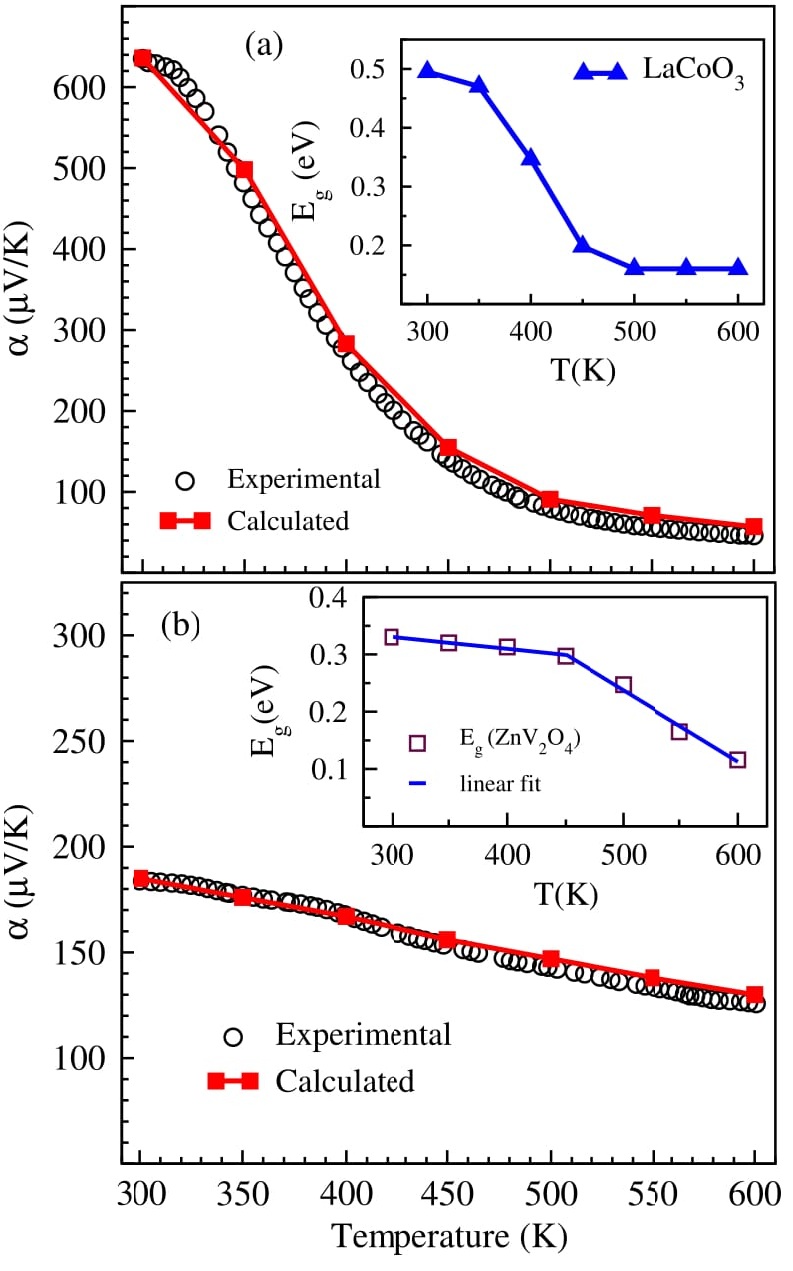}
\end{figure}

\begin{figure}
\caption{}
\includegraphics[scale=0.5]{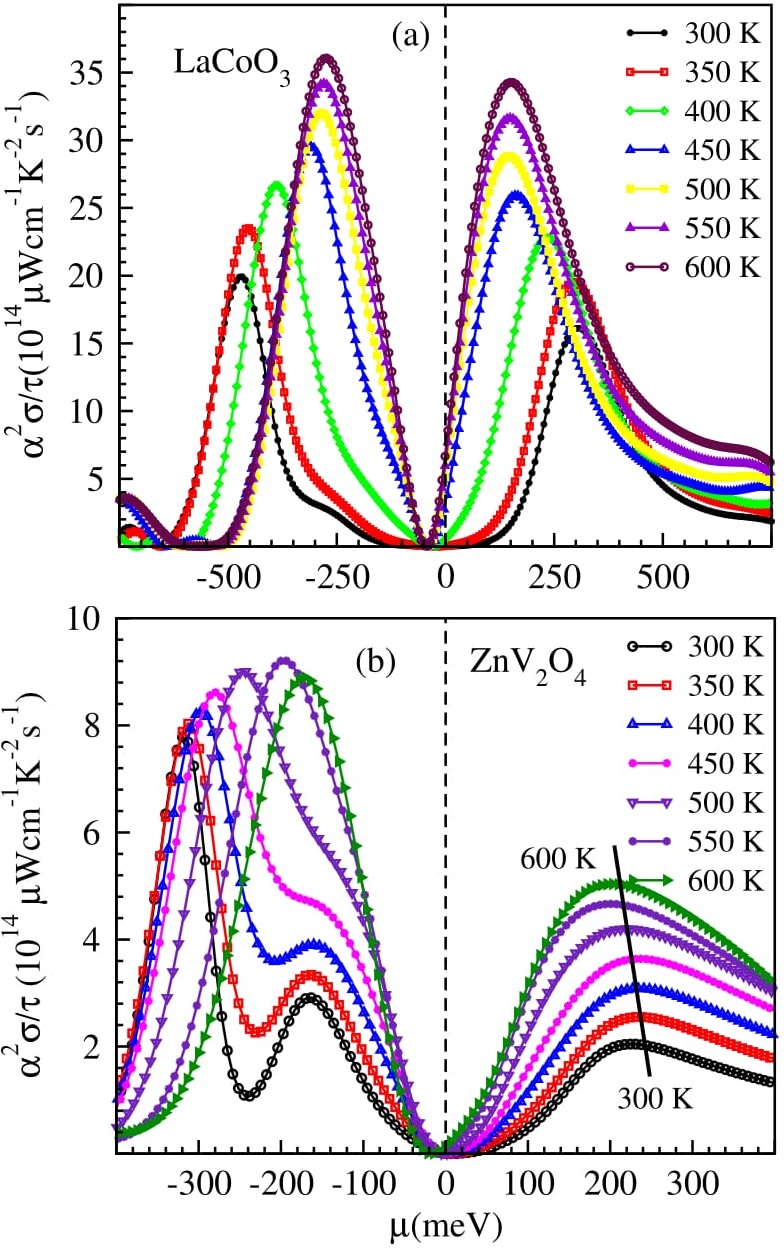}
\end{figure}

\begin{figure}
\caption{}
\includegraphics[scale=0.35]{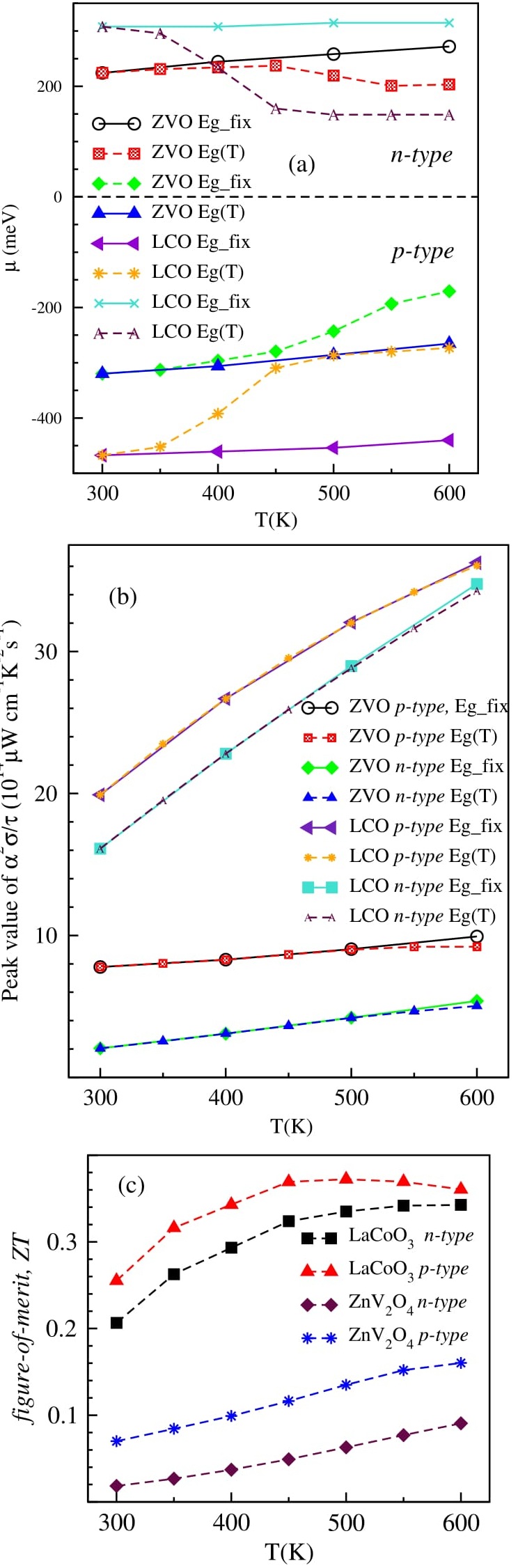}
\end{figure}


\begin{thebibliography}{99}

\bibitem{DiSalvo} F. J. DiSalvo, \textit{Science} \textbf{238}, 703 (1999).
\bibitem{Bell} L. E. Bell, \textit{Science} \textbf{321}, 1457 (2008).
\bibitem{Snyder} G. Jeffrey Snyder and E.S. Toberer, \textit{Nat. Mater.} \textbf{7}, 105 (2008).
\bibitem{Trit} T. M. Tritt, \textit{Semiconductors and Semimetals: Recent Trends in Thermoelectric Materials Research \textbf{I $\&$ II}} (San Diego: Academic, 2001).
\bibitem{Madsen}G.K.H. Madsen and D.J. Singh, \textit{ Comput. Phys. Commun.} \textbf{175}, 67 (2006).
\bibitem{Rowe} D. M. Rowe,\textit{ Thermoelectric Handbook: Micro to Nano} (Boca Raton: CRC Press, Taylor $\&$ Francis Group, 2006).
\bibitem{Sholl} D. S. Sholl and J. A. Steckel, Density Functional Theory: A Practical Introduction (Wiley, Hoboken, N.J., 2009).
\bibitem{singhDJ} D. J. Singh, PlanewaVes, Pseudopotentials and the LAPW Method; Kluwer Academic Publishers: Dordrecht, The Netherlands, 1994.
\bibitem{sonu} S. Sharma and S. K. Pandey, J. Phys. D: Appl. Phys. \textbf{47}, 445303 (2014).
\bibitem{Salameh} S. Ahmad and S. D. Mahanti, Phys. Rev. B \textbf{81}, 165203 (2010).
\bibitem{Singh} S. Singh and S. K. Pandey, arXiv:1606.01539.
\bibitem{SSingh} S. Singh, R. K. maurya and S. K. Pandey, J. Phys. D: Appl. Phys. \textbf{49}, 425601 (2016). 
\bibitem{Saurabh} S. Singh and S. K. Pandey, arXiv:1508.04739.
\bibitem{Blaha}P. Blaha, K. Schwarz, G. K. H. Madsen, D. Kvasnicka and J. Luitz, WIEN2k \textit{An Augmented Plane Wave plus Local Orbitals Program for Calculating Crystal Properties} (Karlheinz Schwarz Technische Universit at Wien, Austria) 2001, ISBN 3-9501031-1-2.
\bibitem{Radaelli}P. G. Radaelli and S. W. Cheong \textit{Phys. Rev. B} \textbf{66} 094408 (2002).
\bibitem{Reehuis} M. Reehuis, A. Krimmel, N. Buttgen, A. Loidl, and A. Prokofiev Eur. Phys. J. B \textbf{35}, 311 (2003).
\bibitem{Combineref} For the detailed information of experimental studies and theoretical calculations performed on these compounds, users are requested to refer the Ref. \textbf{11} and \textbf{12}
\bibitem{Kohn}W. Kohn and L. J. Sham, Phys. Rev. B \textbf{140}, A1133 (1965).
\bibitem{Perdew}J. P. Perdew and Wang Y Phys. Rev. B \textbf{45} 13244 (1992).
\bibitem{Ansimov}V. I. Ansimov, J. Zaanen, and O. K. Andersen, Phys. Rev. B \textbf{44}, 943 (1991).
\bibitem{Hedin} L. Hedin, Phys. Rev. \textbf{139} A796 (1965).
\bibitem{Tran} F. Tran and Peter Blaha, Phys. Rev. Lett. \textbf{102}, 226401 (2009).
\bibitem{Rogers} D. B. Rogers, R. J. Arnott, A. Wold and J. B. Goodenough \textit{J. Phys. Chem. Solids} Pergamon Press, \textbf{24} 347 (1962).
\bibitem{Arima}T. Arima, Y. Tokura, and J. B. Torrance \textit{Phys. Rev. B} \textbf{48} 17006 (1993).
\bibitem{Ashcroft} N.W. Ashcroft, N.D. Mermin, in: D.G. Crane (Ed.), Solid State Physics, vol. \textbf{239}, Saunders College Publishing, New York, (1976).
\bibitem{Varshni} Y. P. Varshni, \textit{Physica} \textbf{34}, 149 (1967).
\bibitem{Ishitsuka} Y. Ishitsuka, T. Ishikawa, R. Koborinai, T. Omura, and T. Katsufuji, Phys. Rev. B \textbf{90}, 224411 (2014).
\bibitem{Pillai}C. G. S. Pillai and A. M. George, \textit{Inter. J. of Therm.} \textbf{4}, 2 (1983).

\end{thebibliography}
\end{document}